\documentclass[12pt]{article}
\usepackage{amssymb}
\usepackage{amsmath}
\usepackage{epsf}
\def\beq{\begin{equation}}
\def\eeq{\end{equation}}
\def\barr{\begin{array}}
\def\earr{\end{array}}

\def\gev{\; {\rm GeV} }
\def\tev{\; {\rm TeV} }
\def\bea{\begin{eqnarray}}
\def\eea{\end{eqnarray}}

\def\bold#1{\setbox0=\hbox{$#1$}
     \kern-.025em\copy0\kern-\wd0 
     \kern.05em\copy0\kern-\wd0 
     \kern-.025em\raise.0433em\box0 }

\def\ie{ {\em i.e.}}

\def\etal{ {\em et al.}}
\def\gev{ {\rm ~GeV}}
\def\tev{ {\rm ~TeV}}

\begin{document}
\vspace*{-1in}
\renewcommand{\thefootnote}{\fnsymbol{footnote}}
\begin{flushright}
MRI-P-011101\\
\texttt {hep-ph/0111083} 
\end{flushright}
\vskip 5pt
\begin{center}
{\Large{\bf Muon anomalous magnetic moment constrains models with excited leptons}}
\vskip 25pt
{\sf Subhendu Rakshit 
\footnote{E-mail address: srakshit@mri.ernet.in}}  
\vskip 10pt 
{\em Harish-Chandra Research Institute, Chhatnag Road, Jhusi,\\ Allahabad
211 019, India }\\

\vskip 20pt

{\bf Abstract}
\end{center}

\noindent
\begin{quotation}
  {\small We explore the possibility of the existence of excited
    leptons in the light of recent data on muon $g - 2$ from BNL. We
    have been able to put stringent bounds on the relevant parameter
    space.
\vskip 10pt
\noindent   
} 
\end{quotation}

\vskip 20pt  

\setcounter{footnote}{0}
\renewcommand{\thefootnote}{\arabic{footnote}}
%
%
\section{Introduction}
The recent measurement of the anomalous magnetic moment ($a_{\mu}$) of
the muon by  the E821 experiment~\cite{BNL} at  BNL seem to indicate a
$2.6\sigma$ deviation  from   theoretical predictions  based  on   the
Standard     Model   (SM),  more     precisely,  $\delta a_{\mu}\equiv
a_{\mu}^{expt} - a_{\mu}^{SM} = 426(165) \times 10^{-11}$. The ongoing
analysis of collected data and accumulation of fresh data is likely to
reduce the  errors even further.  The required additional contribution
to explain this anomaly is nearly three times the standard electroweak
contribution. This deviation might  be indicative of some new  physics
beyond  the Standard Model (SM)  and this has  led to consideration of
several  extensions of the SM~\cite{haber}. Of  course one should keep
in   mind  that  the   existing SM   calculations   for the   hadronic
contributions to  $a_{\mu}$ are questionable~\cite{hadronic} and there
is no universal consensus   amongst different groups  performing these
calculations. But this does not undermine the credibility of exploring
new physics options beyond the SM.

In  the SM  leptons and quarks   are treated as fundamental point-like
objects.    But  ordinary leptons    might  be made  up  of  some more
fundamental   particles~\cite{review}  interacting  in  a  very strong
confining interaction characterized by  some scale $\Lambda \sim {\cal
  O}(1)\tev$.     A    possible   prediction     of  these   composite
models~\cite{composite} is the existence of excited states of ordinary
leptons. It was pointed out in  Ref.~\cite{renard2} that these excited
leptons  can give rise  to a large magnetic   moment for the muon. The
reported  anomaly  on $a_{\mu}$ can  be  explained  in the presence of
excited  leptons and this, in  turn,   severely restricts the  allowed
parameter space.

\section{Description of the model}
We consider excited state of  the muon ($\mu^*$) and its corresponding
excited   neutrino ($\nu_{\mu}^{*}$) to   contribute  at  one loop  to
$a_{\mu}$~\cite{others}.  In this study we  shall confine ourselves to
the lowest lying excited states of spin and isospin $1/2$, although in
principle, other  excited states with higher  spin and various isospin
values might also  exist~\cite{isospin}.  However, this is  sufficient
for  a {\em conservative order of  magnitude estimation} as  we do not
see any  compelling   reason  for any fine   cancellation  between the
contributions from the first excited state and other excited states.

These excited  muons are  very massive  in comparison to  the ordinary
ones. To motivate this  large mass gap,  we assume that these  excited
states  acquire their masses  prior to $SU(2)\times U(1)$ breaking and
so  both of   their  left- and right-handed    states will be  in weak
isodoublets. As  a   consequence,  the  interactions of   $\mu^*$  and
$\nu_{\mu}^{*}$ with the gauge bosons are vector-like~\cite{boudjema}.
The corresponding Lagrangian can be written as,
\begin{equation}
{\cal L}_{VF'F} = - \sum_{V=\gamma, Z, W} C_{VF'F} \overline{F'} 
                \gamma_{\mu} F V^{\mu} + \makebox{h.c.}, 
\end{equation}
where $F,F' = \mu^*, \nu_{\mu}^{*}$. The constants $C_{VF'F}$'s are given by  
\begin{xalignat}{2}
C_{Z \mu^*\mu^*} &= \frac{g}{2\cos\theta_W}(1-2\sin^2\theta_W), & 
C_{Z \nu_\mu^*\nu_\mu^*} &= - \frac{g}{2\cos\theta_W} \nonumber\\ 
C_{\gamma\mu^*\mu^*} &=  e  & C_{W \nu_\mu^*\mu^*}&= - \frac{g}{\sqrt 2}\nonumber
\end{xalignat}
In the above Lagrangian  and in the  following ones, we  will indicate
only those interactions relevant for our subsequent discussions.

On  the other hand, the Lagrangian   describing the transition between
the ordinary muon and the excited muon may look like:
\begin{equation}
\begin{split}
{\cal L}_{Ff} &= 
\frac{1}{4\Lambda} \;{\mathbf{\overline{F}_R}} \; \sigma^{\mu\nu}\,
          \left[ 
                 a\, {\boldsymbol\tau} \cdot \,{\mathbf{W}}_{\mu\nu}
                +b\, Y \, B_{\mu\nu}
          \right]{\mathbf{f_L}}  \\ 
               &+ 
\frac{1}{4\Lambda^2} \;{\mathbf{\overline{F}_L}} \; \sigma^{\mu\nu}\,
          \left[ 
                 a'\, {\boldsymbol\tau}\cdot \,{\mathbf{W}}_{\mu\nu}
                +b'\, {Y} \, B_{\mu\nu}
          \right]{{f_R}} \, \mathbf{\Phi} + \, \text{h.c.}\,.  \label{master}
\end{split}
\end{equation}
It is customary  to parametrize  the new  coupling  parameters as  $a=
g\,f_L$, $a'= g'\,f'_L$,   $b=  g\,f_R$ and  $b'=   g'\,f'_R$.   Here,
\{$f_{_L}$,  $f_{_R}$\}  and  \{$f'_{_L}$, $f'_{_R}$\}  are the weight
factors   associated with    the  gauge  groups   $SU(2)$  and  $U(1)$
respectively and  they arise from  the underlying  dynamics describing
the  compositeness.   However,   in   this  paper, for   the   sake of
convenience, we shall be using the  coupling parameters $a$, $b$, $a'$
and $b'$ instead of $f_{_L}$, $f_{_R}$, $f'_{_L}$ and $f'_{_R}$.

In the above Lagrangian,
\begin{alignat}{4}
{\mathbf{F_L}} = {\begin{pmatrix} \nu_\mu^{*} \\ \mu^* \end{pmatrix}}_L, \qquad &  
{\mathbf{F_R}} = {\begin{pmatrix}\nu_\mu^{*} \\ \mu^* \end{pmatrix}}_R, \qquad & 
{\mathbf{f_L}} = {\begin{pmatrix}\nu_\mu \\ \mu \end{pmatrix}}_L,\qquad &
{{f_R}} = \mu_{R}.\nonumber
\end{alignat}
$\mathbf{\Phi}$ is the SM Higgs doublet.  

It is important to note in ${\cal  L}_{Ff}$ that we have included both
left-  and  right-handed    excited leptons.   It   was observed    in
Ref.~\cite{renard2} that this  might lead to  a  large magnetic moment
for  the ordinary leptons. To  avoid this, it is  a common practice to
consider only the right-handed excited  fermions interacting this way. 
Here we allow  both type of couplings to  review the bounds imposed on
the coupling parameters in view of the  recent anomaly in the magnetic
moment of the  muon. Inclusion of  left-handed excited leptons require
consideration of operators of  higher  dimensionality than  those  for
right-handed ones.  So  the interactions arising   from second line of
eqn.~\ref{master}  will suffer   an  additional suppession  factor  of
$v/\Lambda$  in comparison  to those from   the first line.  Here  $v$
stands for the VEV of the SM Higgs.

It is also possible to include a term like 
\begin{equation}
\frac{1}{4\Lambda^2} \;{\mathbf{\overline{F}_L}} \; \sigma^{\mu\nu}\,
          \left[ 
                 a''\, {\boldsymbol\tau}\cdot \,{\mathbf{W}}_{\mu\nu}
                +b''\, {Y} \, B_{\mu\nu}
          \right]{{\nu_R}} \, \mathbf{\widetilde\Phi} + \, \text{h.c.}
\end{equation}
in eqn.~\ref{master}, which needs introduction of two more couplings
$a''$ and $b''$. For the sake of minimality, we set them to zero. In
the above $\mathbf{\widetilde\Phi} \equiv -i \sigma_2
\mathbf{\Phi}^*$, $\sigma_2$ being the second Pauli matrix. 

The interaction Lagrangian for $\mu$ and $\nu_{\mu}^*$ with a single gauge
boson obtained from eqn.~\ref{master} looks like:
\begin{equation}
{\cal L}_{VFf} = - \frac{1}{2\Lambda} \sum_{\scriptscriptstyle V=\gamma, Z, W}  
                \overline {F} 
                \sigma_{\mu\nu}\,(D_{VFf}^L P_L + D_{VFf}^R P_R) f \,
                \partial^{\mu}\,V^{\nu} + \makebox{h.c.}, 
\end{equation}
where $f = \mu, \nu_{\mu}$. The constants $D_{VFf}$'s are given by,  
\begin{xalignat}{2}
D_{Z \mu^*\mu}^L &= -a \cos\theta_W + b \sin\theta_W \nonumber & 
D_{Z \mu^*\mu}^R &= \frac{v}{\Lambda}(-a' \cos\theta_W + b' \sin\theta_W)\\ 
D_{\gamma \nu_\mu^*\nu_\mu}^L &= ~~\,a \sin\theta_W - b \cos\theta_W & 
D_{\gamma \nu_\mu^*\nu_\mu}^R &= 0 \nonumber\\
D_{\gamma \mu^*\mu}^L &= -a \sin\theta_W - b \cos\theta_W & 
D_{\gamma \mu^*\mu}^R &= \frac{v}{\Lambda} (-a' \sin\theta_W - b' \cos\theta_W) \nonumber\\
D_{W \nu_\mu^* \mu}^L &= D_{W \mu^* \nu_\mu}^L = a \sqrt 2  & 
D_{W \nu_\mu^* \mu}^R &= \frac{v}{\Lambda} a \sqrt 2~; ~~~D_{W \mu^* \nu_\mu}^R = 0\nonumber
\end{xalignat}

The Lagrangian describing quartic interactions between $\mu$ and
$\mu^*$ with $W$ and $\gamma$ arising from the non-abelian part of
$W_{\mu\nu}$ in eqn.~\ref{master} can be written as,
\begin{equation}
{\cal L}_{VVFf} = - \frac{1}{2\Lambda}\,\overline {F} 
                \sigma_{\mu\nu} (E_{VVFf}^L P_L + E_{VVFf}^R P_R) f
                 W^{\nu} + \makebox{h.c.},  
\end{equation}
where the constants $E_{VVFf}$'s are given by  
\begin{xalignat}{2}
E_{\gamma W \mu^*\nu_\mu}^L = -E_{\gamma W \nu_\mu^*\mu}^L = -a\;e \sqrt 2; \qquad & 
E_{\gamma W \mu^*\nu_\mu}^R =  0~;
~~~E_{\gamma W \nu_\mu^*\mu}^R = \frac{v}{\Lambda} a'\;e \sqrt 2\nonumber
\end{xalignat}

For the sake of notational convenience, we also define the required SM
interactions in the following Lagrangian:
\begin{equation}
{\cal L}_{Vf'f} = -  \sum_{\scriptscriptstyle V=\gamma, Z, W} 
                  \bar f' \gamma^{\mu} 
                  \left(
                  F^L_{Vf'f} P_L + F^R_{Vf'f} P_R
                  \right) f.
\end{equation}

\section{Excited lepton contribution to $a_\mu$}

With the interactions defined in the previous section, we can calculate
the one-loop diagrams containing excited leptons
\begin{figure}[p]
\vspace*{-3.0cm}
\hspace*{-1.0cm}
\centerline{
\epsfxsize=17.0cm\epsfysize=22.5cm
\epsfbox{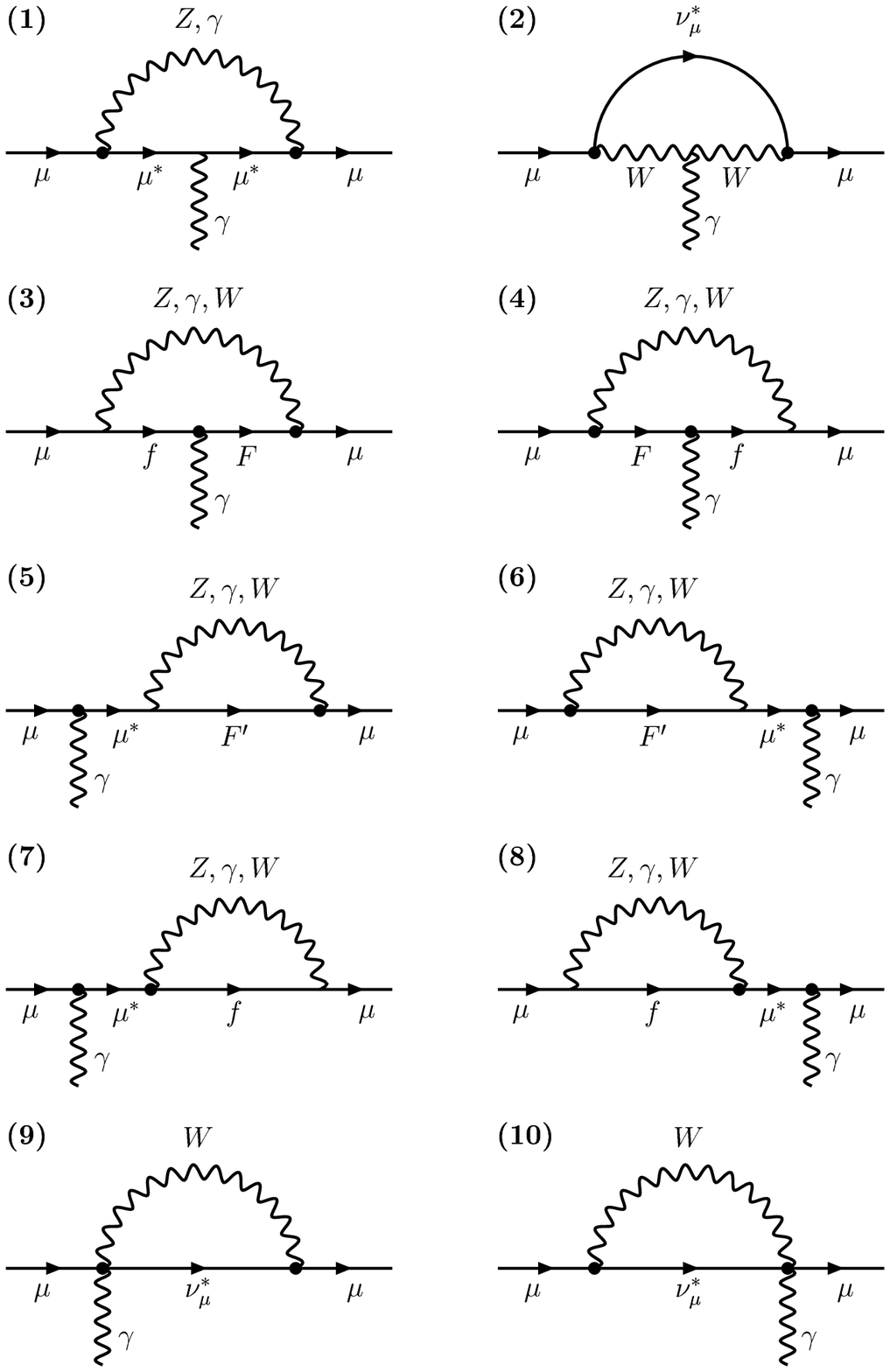}
           }
\vspace*{-3.0cm}
\caption{\em Diagram sets containing excited leptons contributing to $a_{\mu}$} 
      \label{diag}
\end{figure}
contributing to $a_{\mu}$.   It has to be pointed  out that in the SM,
the self-energy type  diagrams do not contribute  to  $a_\mu$ but here
diagrams 5---8 (see Fig.~\ref{diag}) contribute  due to the transition
magnetic  interactions between ordinary   and  excited leptons. It  is
obvious from the Dirac structure of the vertices that the longitudinal
parts of the gauge bosons do not contribute.  As a consequence, in the
absence of  form-factors (see the   following discussion), one has  to
encounter at most quadratic divergences.

There   are several  alternatives   to  make   the momentum  integrals
convergent. One of the options is to apply a  momentum cut-off at $k^2
=     \Lambda^2$~\cite{renard2}.   Another       option  is     to use
form-factors~\cite{renard1}, as one can no longer treat the leptons as
point   particles. We shall    follow this alternative.  We  associate
dipole   form-factors $\left(\frac{1}{1-k^2/\Lambda^2}\right)^2$  with
the  vertices which involve  magnetic  transitions between the excited
leptons with  the ordinary ones. Here $k$  denotes the momentum of the
associated off-shell gauge  boson. Muon magnetic moment calculation in
the  dimensional  regularization  method  was  performed  for  excited
leptons   (with a       different    set  of     interactions)      in
Ref.~\cite{gonzalez}. We make a remark in passing that for an order of
magnitude estimation, the numerical differences coming from the choice
of different regularization schemes are not important~\cite{renard2}.

In the appendix we present analytical expressions of the contributions
to $a_\mu$ from the dominant  diagram sets. The contributions from the
diagrams   9  and  10 containing  the   quartic  coupling   are always
insignificant. Contribution from  diagrams   5---8 can be  more   than
$90\%$ of  the total excited lepton  contribution to $a_\mu$. When $a$
and  $a'$ both are  zero, the $W$  diagrams  do not  contribute as the
$SU(2)$  sector does    not participate  in   the transition  magnetic
interaction  between the ordinary   and the excited leptons.  From the
symmetry  of the diagrams it is  clear that diagram  sets 3 and 4 give
equal contributions.  The same is true for   the other sets  \{5, 6\},
\{7, 8\} and \{9, 10\}.

In this  model there  are  several extra  parameters than the standard
model: (a) the coupling parameters  $a$, $b$, $a'$  and $b'$, (b)  the
masses of $\mu^*$ and  $\nu_{\mu}^{*}$ and (c) the compositeness scale
--- $\Lambda$.  We take  $\mu^*$ and $\nu_{\mu}^{*}$ to be  degenerate
in  mass  ($M_F$),  so that  they  do   not contribute  to  the $\rho$
parameter.

\section{Results and discussion}

In this    section we  shall  mainly  divide  our attention   onto two
categories of couplings.   At first we  shall deal  with the situation
when only one of the coupling constants is  nonzero. In this situation
the  transition magnetic  interactions  between  the ordinary  and the
excited leptons will respect a chiral symmetry,\ie,  only left- or the
right-handed excited lepton  will be coupled  to ordinary fermions. We
shall refer  to this as chirality  conserving interactions.  The other
obviously will  be chirality violating  and will arise when either $a$
and $a'$ or  $b$ and $b'$  will be simultaneously  non-zero.  Magnetic
moment  being  a chirality  violating  operator   one need to  have  a
left-handed muon at one end of the Feynman  diagram and a right-handed
muon  on  the  other. Now  in each  diagram  there are  two transition
magnetic interactions. So to have a  muon magnetic moment one needs to
have  at  least  one mass  insertion on  the  fermion  line.  For  the
chirality violating interactions, one can have a mass insertion on the
excited  lepton  line making   the  contribution large.   But for  the
chirality conserving case, one gets  a helicity flip  on the muon line
making the contribution more suppressed. We can now proceed to discuss
the parameter dependence of $\delta a_\mu$.

\subsection{Dependence on coupling constants}

Now let us discuss the  parameter dependence of the total contribution
to $a_\mu$ from these diagrams. If we consider only $a$ to be nonzero,
then $a_\mu$ increases  quadratically  with it,\ie, $\delta a_\mu  = y
(a/\Lambda)^2$.   For $M_F = 400\gev$    and $\Lambda = 1\tev$,  $y(in
\gev^2) =  1.6~10^{-3}$. For $b$, $a'$ and  $b'$, $y$  is evaluated as
$5.8~10^{-4}$, $4.3~10^{-5}$ and $5.5~10^{-5}$ respectively.

For chirality violating case, it  turns out that the terms  containing
the products $aa'$  or $bb'$ give  the dominant contribution due to  a
mass insertion on the excited  lepton line. Now  for $a= a' v/\Lambda$
or $b=b' v/\Lambda$, the magnetic  transition between the ordinary and
excited leptons are pure vectorial, whereas for $a= - a' v/\Lambda$ or
$b=-  b' v/\Lambda$, the  interactions are axial-vector-like. We shall
deal with these two extremes and in all the above cases the dependence
of $\delta  a_\mu$  on $a$ or  $b$ is  obviously  quadratic as before. 
$y=9.9$ for $a=   a'  v/\Lambda$  and $y=5.3$ for   $b=b'  v/\Lambda$. 
Comparing these values of $y$ with the  previous case it is clear that
in this case there will be a large contribution from excited leptons.

\subsection{Dependence on $M_F$}

\begin{figure}[htp]
\vspace*{-3.0cm}
\hspace*{-0.5cm}
\centerline{
\epsfxsize=7.0cm\epsfysize=9.0cm
\epsfbox{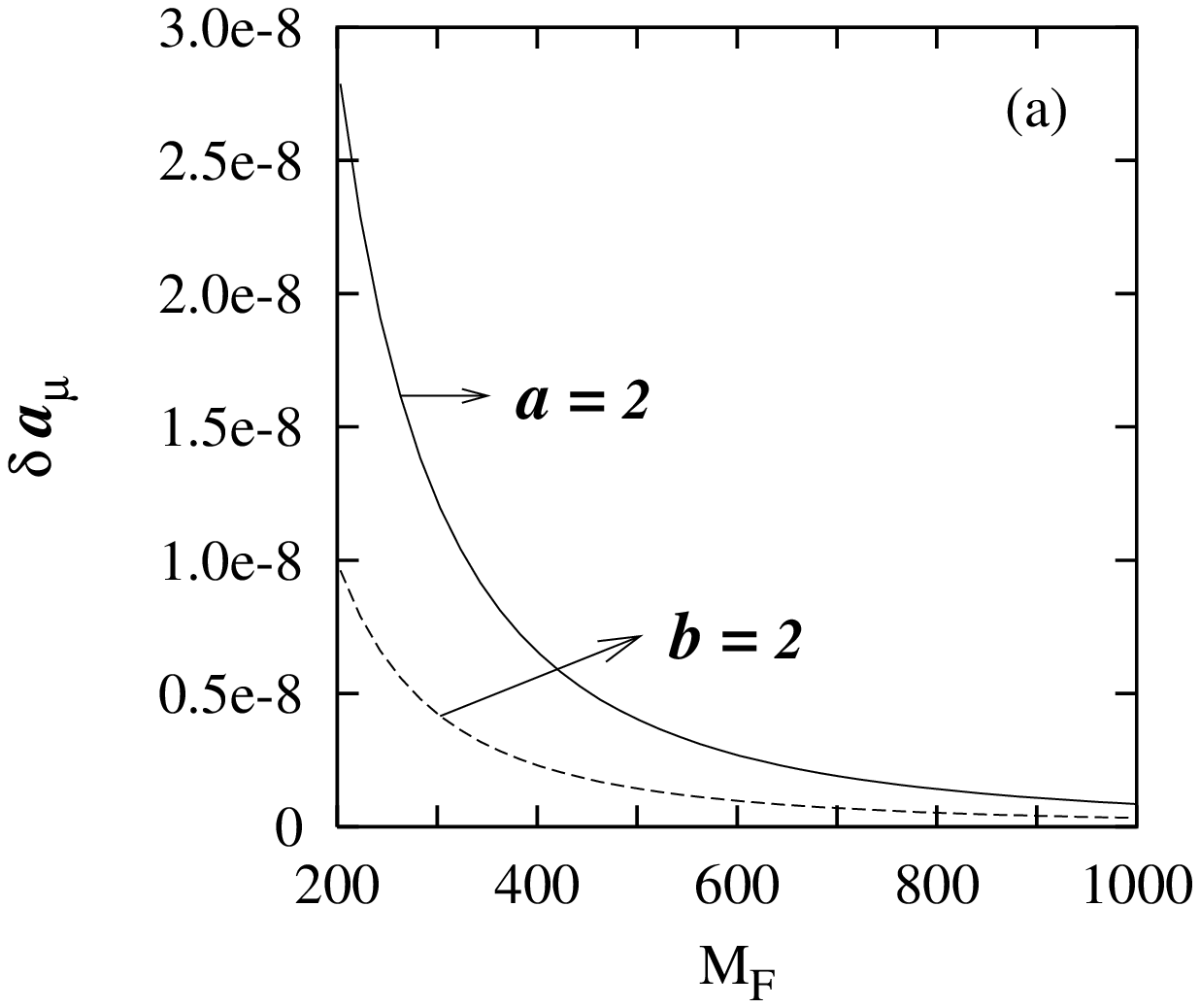}
\vspace*{-0.0cm}
\hspace*{-0.cm}
\epsfxsize=7.0cm\epsfysize=9.0cm
\epsfbox{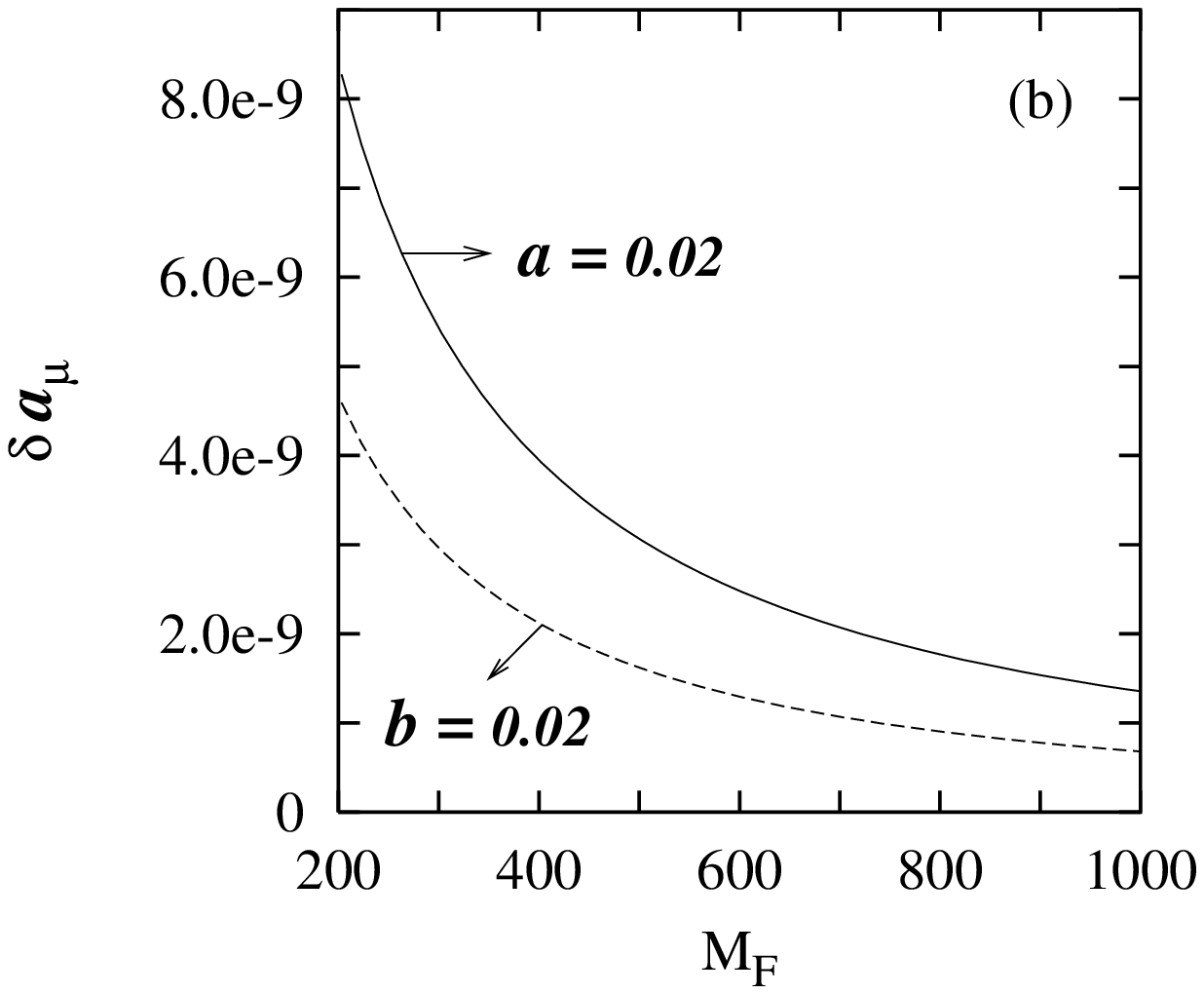}
\vspace*{-2.0cm}
}                

\caption{\em 
  Dependence of $a_{\mu}$ on excited lepton mass. In (a) we have taken
  either $a$ or $b$  to  be non-zero. In   (b) we have considered  the
  chirality violating cases when $a= a'v/\Lambda$ or $b= b'v/\Lambda$.
  We have put $\Lambda = 1\tev$.}
      \label{mfdep}
\end{figure}

For chirality conserving  case, $\delta a_\mu$  is almost proportional
to $1/M_F^2$ (see Fig.~\ref{mfdep}(a)). The reason  behind is that the
diagram  sets   5---8  give  the  dominant  contributions.   In  these
diagrams, we  can  pick $1/M_F^2$  from   the propagator  joining  the
one-loop part to the external photon vertex. As discussed earlier, one
can have a helicity flip on the muon line only.  So no other factor of
$M_F$  comes   into the  picture.   $M_F$   dependence coming from the
propagators inside the loops makes a  little deviation from $1/M_F^2$. 
But for chirality violating  case, there is  another power of $M_F$ in
the numerator due to  a helicity flip.   Hence, then the dependence is
more like $1/M_F$ (see Fig.~\ref{mfdep}(b)).

\subsection{Dependence on $\Lambda$}

\begin{figure}[hbp]
\vspace*{-3.0cm}
\hspace*{-0.5cm}
\centerline{
\epsfxsize=7.0cm\epsfysize=9.0cm
\epsfbox{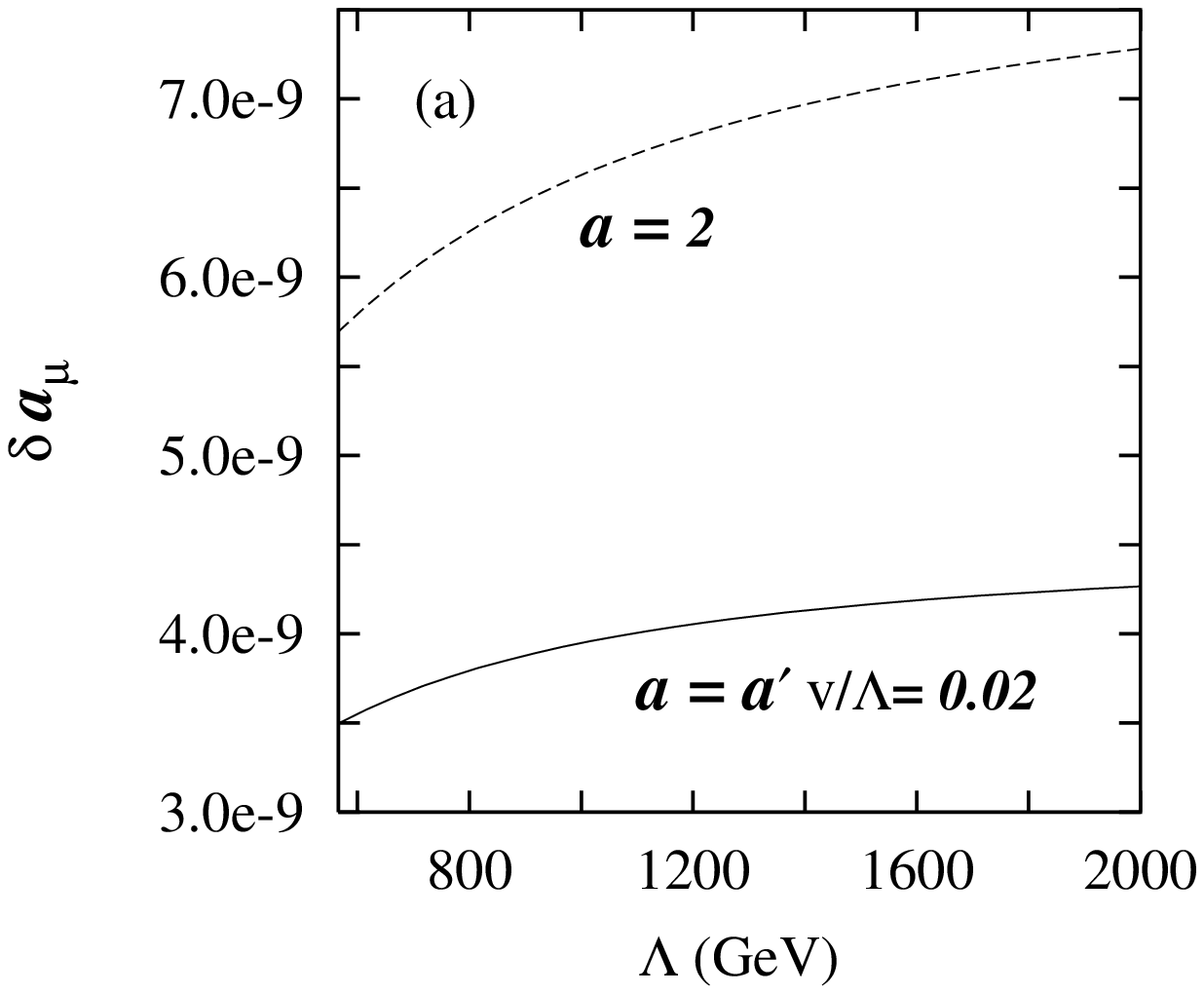}
\vspace*{-0.0cm}
\hspace*{-0.cm}
\epsfxsize=7.0cm\epsfysize=9.0cm
\epsfbox{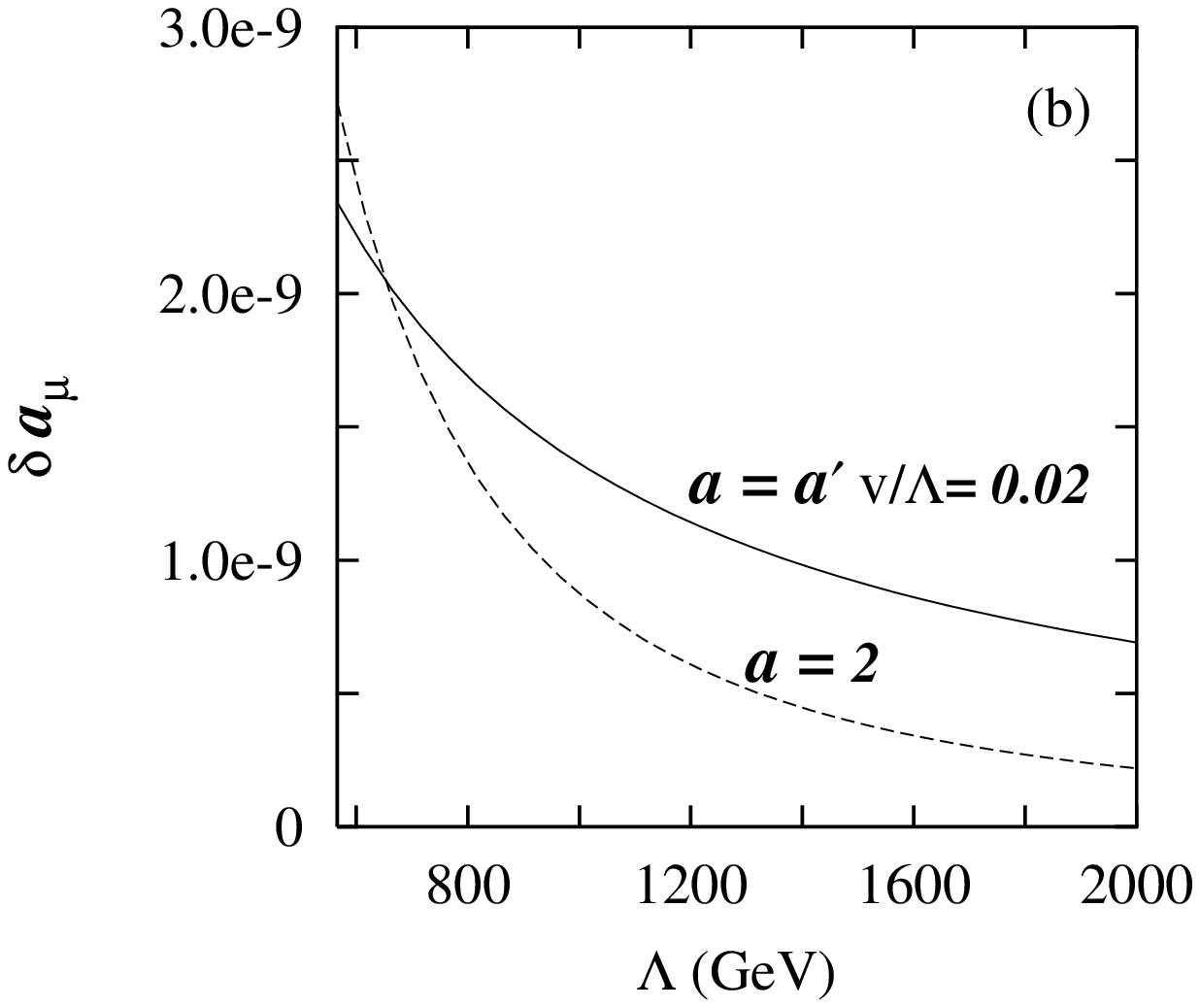}
\vspace*{-2.0cm}
}                
\caption{\em Dependence of  $\delta a_{\mu}$ on the compositeness 
  scale $\Lambda$. In diagram (a) we have kept $M_F$ fixed at
  $400\gev$. In (b) we take $M_F \sim \Lambda$.} \label{Ldep}
\end{figure}

In Fig.~\ref{Ldep}, we explore the dependence of $\delta a_\mu$ on the
compositeness scale    $\Lambda$.  In fig.   (a)    we  see that   the
contribution increases and then rather than blowing up saturates as we
keep  on  increasing  $\Lambda$.  Here we have  kept    $M_F$ fixed at
$400\gev$. But  if we take $M_F \sim  \Lambda$, then it  falls rapidly
(see fig. (b)). The reason  is that in this  case the $M_F$ dependence
also adds  up to the aforesaid  $\Lambda$ dependence.  With respect to
this situation, the plotted behaviour in (a) is rather flat.

\subsection{Constraints on the parameter space}

\begin{figure}[htb]
\vspace*{-2.0cm}
\hspace*{-0.5cm}
\centerline{
\epsfxsize=7.0cm\epsfysize=9.0cm
\epsfbox{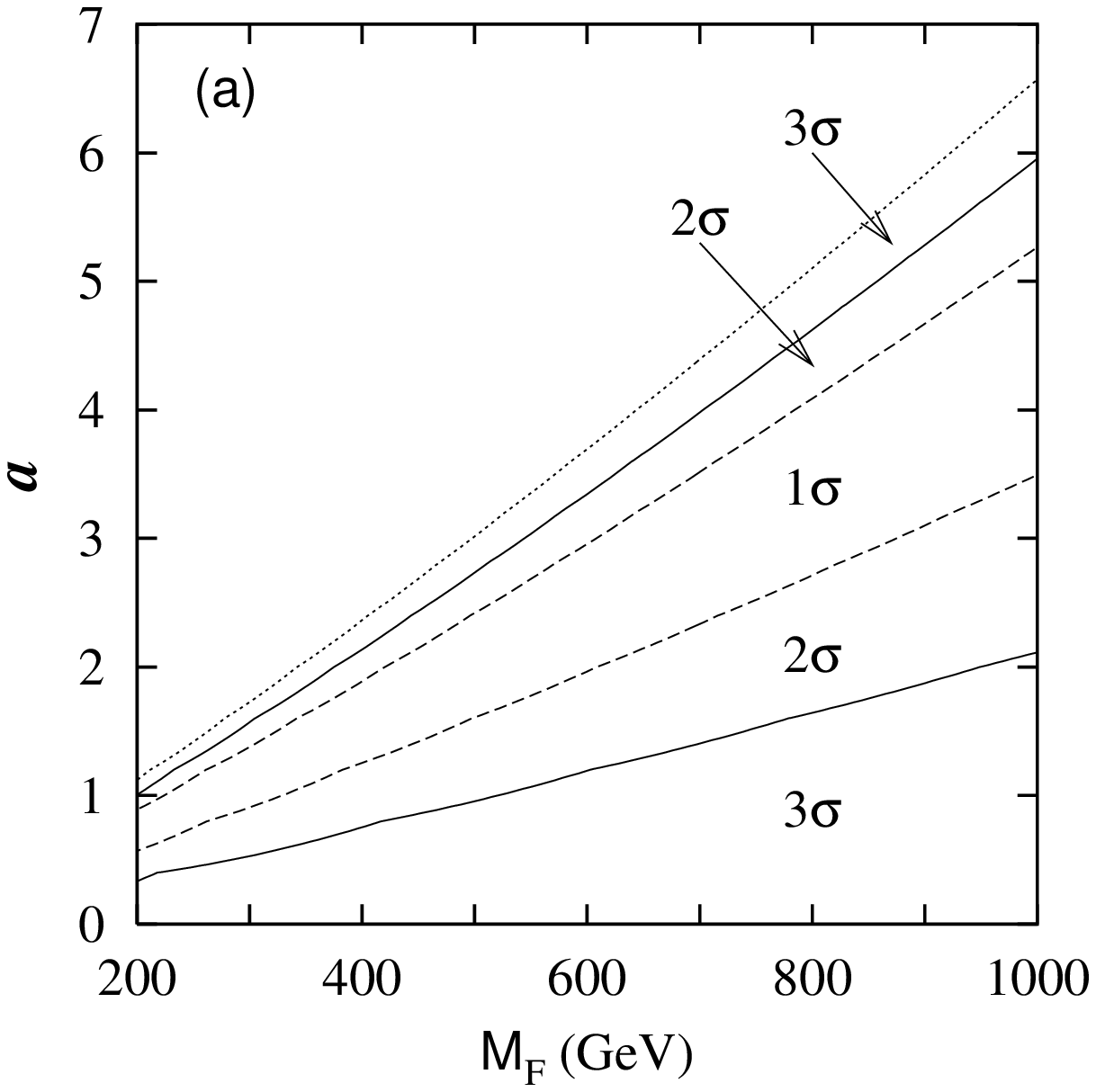}
\vspace*{-0.0cm}
\hspace*{-0.cm}
\epsfxsize=7.0cm\epsfysize=9.0cm
\epsfbox{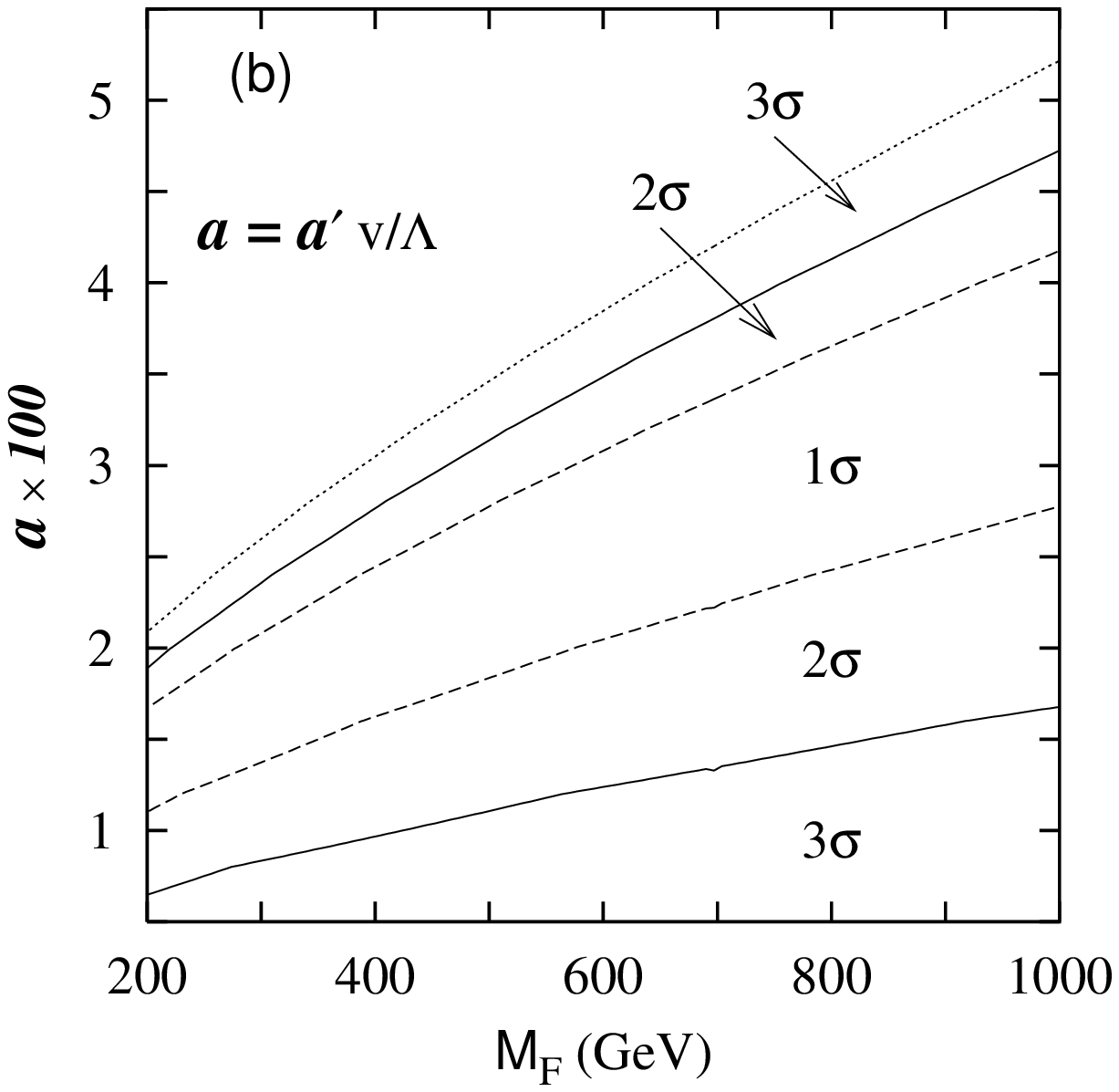}
\vspace*{-3.0cm}}
\caption{\em Contour plots for fixed values of 
  $\delta a_{\mu}$ in $M_F$ Vs. coupling constant plane.}
      \label{cont}
\end{figure}

To  get  an   idea of  the  parameter  space  allowed by    the recent
measurement on $a_\mu$, it is convenient to look  at the contour plots
for $\delta a_\mu$  in the coupling constant ---  $M_F$  plane. Let us
discuss the situation when only one of the couplings is non-zero.  For
simplicity of presentation,  we take only $a$  to be non-zero.  Now we
have seen that  in this  case $\delta  a_\mu  \propto a^2$. The  $M_F$
dependence   is  more like  $1/M_F^2$.    So we  expect  almost linear
behaviour in  the contours $a  \sim  z_1 \, \sqrt{\delta a_\mu}  M_F$. 
$z_1  (in \gev^{-1})$   is  evaluated (for  $\Lambda=1\tev$)  as $66$,
$108$, $450$ and $342$  for $a$, $b$, $a'$  and $b'$ respectively. For
an  order of magnitude  estimation this formula   works quite well. In
Fig.~\ref{cont}(a), we   indicate regions in  the  $a$ --- $M_F$ plane
allowed    by the  recent   data   on   $g-2$ measurement with   their
corresponding confidence levels.

Now let us  deal with two non-zero  coupling constants. We shall start
with   the situation  when the  excited  leptons  have  pure vectorial
transition magnetic coupling with the  ordinary leptons. In this  case
the contribution will be positive. Argument in the same line as before
leads to  a relation $a \sim z_2  \, \sqrt{\delta a_\mu \, M_F}$. $z_2
(in  \gev^{-1/2})$ is given by  $16.5$ and  $22.9$ for $a=a'v/\Lambda$
and $b=b'v/\Lambda$ respectively.

In    Fig.~\ref{cont}(b),    we  present    the    contour   plot  for
$a=a'v/\Lambda$.  We  see that  the constraints  arising out  in  this
situation are more  stringent.  If one  of them is negative, the total
contribution is also negative which is allowed  at the most at a level
of $3\sigma$.

\section{Summary and conclusions}

We  have   discussed both  types   of transition  magnetic moment-type
coupling  of excited  leptons  to ordinary  leptons  -- (a)  chirality
conserving and  (b)   chirality violating.  For  the   later type   of
coupling, one  gets a huge  contribution to  $a_{\mu}$~\cite{renard2}. 
We have explored how these models with excited leptons can explain the
recent anomaly on muon   anomalous magnetic moment   measurement. This
puts stringent constraints on the relevant parameter space.

Excited leptons can also contribute radiatively in  the muon mass. Any
such correction to the muon mass can be absorbed in the bare muon mass
using on-shell renormalization conditions.  In the chirality violating
case (which is not  a natural  theory),  one can radiatively  generate
muon mass this way even taking tree level muon mass  to be zero. There
the contribution will be large due to a factor of  $M_F$ coming from a
helicity flip on the  muon line and it  will not be possible to absorb
it in the bare   mass~\cite{marciano1}, which is zero.   Consequently,
one can  get stringent bounds on the  coupling  parameters for a given
compositeness scale $\Lambda  \sim {\cal O}(1)\tev$. However,  here we
do not   try to generate muon  mass  in this  way.    In the chirality
violating case, the  weight factors --- $f_i$'s  turn  out to be  very
small and this has to be ensured by the underlying dynamics describing
compositeness.

We find  that for the chirality violating  scenario, the values of the
parameters $a$  and  $a'$ (or $b$ and  $b'$)  with a relative negative
sign are  disfavoured  by the data. The   constraints we get are  more
stringent than the  limits  obtained from direct searches  of  excited
leptons  at  LEP, where   for   unit transition  magnetic  moment-type
couplings,  the lower  bound   on   excited  lepton mass  is    ${\cal
  O}(100)\gev$~\cite{vachon}.   This rules  out  any hope  for finding
these particles via direct  searches, if they have chirality violating
couplings.    On the other  hand,   the scenario for particles  having
chirality conserving  couplings may not  be that grim, but  the recent
BNL measurement    of $a_\mu$  leaves   behind  a  very  small allowed
parameter space. Recently   search prospects for  excited leptons have
been studied for the  Fermilab Tevatron~\cite{tevatron} and CERN Large
Hadron Collider~\cite{LHC}.  The studies have been  made for $e^*$ and
$\nu_e^*$  in the chirality  conserving scenario.  However, if we take
the electron  and muon detection  efficiencies to be similar  then the
same bounds can be realized for excited muons as well.  In the case of
the  Tevatron     (run   II),    it  has    been   claimed  that   for
$f_L/\Lambda>1\tev^{-1}$, an  excited lepton mass  less than $250\gev$
can  be ruled out at $95\%$  CL. For LHC, the bounds  will be far more
stringent.  It will be  capable of ruling out  a region (at $95\%$ CL)
$f_L/\Lambda>0.7\tev^{-1}$  for  excited lepton  masses less than than
$1\tev$ and will   be sensitive upto $2\tev$.  Hence,  we see  that if
excited leptons are the culprit for the recent  muon anomaly, they can
leave  their footprints in the Tevatron  or in the   LHC, if they have
chirality conserving  transition  magnetic coupling  with the ordinary
leptons. But even if these colliders fail to  detect them, the excited
leptons  can   still explain  the  muon  anomaly   if  they have small
chirality violating couplings.

\vskip 20pt
\noindent{\large\bf Acknowledgements}\\\\
S.R. would like to thank Debajyoti Choudhury for suggesting this work,
help in algebraic computation and for numerous comments and
discussions. He thanks Biswarup Mukhopadhyaya, Uma Mahanta, Debrupa
Chakraverty and Anindya Datta for illuminating discussions. He also
thanks the Abdus Salam International Centre for Theoretical Physics,
Italy for hospitality where a major part of this work was carried out.

\newpage
\noindent{\Large\bf Appendix}\\
\vskip 10pt

The total contribution of excited leptons to anomalous magnetic moment
of muon is given by,
\begin{equation}
\delta a_{\mu} = \frac{m_\mu}{16\pi^2}\, \Lambda^2 \, 
                           \left[
                           \displaystyle\int_{0}^1 dx_1 
                           \displaystyle\int_{0}^{1-x_1} dx_2 
                           \displaystyle\int_{0}^{1-x_1-x_2} dx_3 
                           \sum_{i=1}^{4} X_i
                           +
                           \displaystyle\int_{0}^1 dx_1 
                           \displaystyle\int_{0}^{1-x_1} dx_2 
                           \sum_{i=5}^{8} X_i
                           \right] \nonumber
\end{equation}
The $X_i$'s are contributions from different diagram sets (Fig.~\ref{diag}).
\begin{align}
\begin{split}
X_1 &= 8 \Lambda^4 x_1^3   
  \left( \frac{ f_{1b}  {M_F} \left( 1 - 9  {x_1} - 9  {x_2}\right) }
              { {\Delta_1}^4}  \right . \\
 &\quad - \left . 
     {m_{\mu}} \left( 1- {x_1} - {x_2} \right)  f_{1a} 
     \left( - \frac{ 8 {{M_F}}^2 
          \left(1 - {x_1} -  {x_2} \right) }
          { { {\Delta_1}}^5} + 
       \frac{
          \left( 5 + 4  {x_1} + 4  {x_2} \right) }
           {{ {\Delta_1}}^4} 
      \right)  
  \right)  \nonumber
\end{split}\\
X_2 &= - 8 \Lambda^4 (1-x_1-x_2-x_3)^3\frac{\left( 3  f_{2b}  {M_F} 
       \left( 3  {x_2} - 2 \right)  - 
       f_{2a}  {m_{\mu}}  {x_2} 
       \left( 13 - 12  {x_2} \right)  \right) }
    {{ {\Delta_2}}^4} \nonumber
\\
\begin{split}
X_3 &= 24 {x_1} 
  \left( - 
    \frac{2  f_{3c}  {M_F}}
     {3 { {\Delta_3}}^2} - \frac{2  f_{3d}  {m_f} 
        {M_F}  {m_{\mu}}}{3 
       { {\Delta_3}}^3} (1- x_1 - x_3) +  
    \frac{ f_{3a}  {m_{\mu}}  }
     {{ {\Delta_3}}^2} (x_1 + x_3) \right) \nonumber
\end{split}
\\
\begin{split}
X_4 &= 24 {x_1} 
  \left( - 
    \frac{2  f_{4b}  {M_F}}
     {3 { {\Delta_4}}^2} - 
    \frac{2  f_{4d}  {m_f} 
        {M_F}  {m_{\mu}} }{3 
       { {\Delta_4}}^3} (1 - x_1 - x_3) + 
    \frac{ f_{4a}  {m_{\mu}} }
     {{ {\Delta_4}}^2} (x_1 + x_3)  \right) \nonumber
\end{split}
\\
\begin{split}
X_5 &= 24 {x_1} \left( \frac{2  f_{5b} 
        {M_F}}{ {\Delta_5}} + 
    \frac{2  f_{5a}  {m_{\mu}}}
     { {\Delta_5}} + 
    \frac{f_{5a}  {M_F} 
        {M_{F'}}  {m_{\mu}}  {x_2}}
       {{ {\Delta_5}}^2} \right) \nonumber
\end{split}
\\
\begin{split}
X_6 &= 24 {x_1} \left( \frac{2  f_{6b} 
        {M_F}}{ {\Delta_6}} + 
    \frac{2  f_{6a}  {m_{\mu}}}
     { {\Delta_6}} + 
    \frac{  f_{6a}  {M_F} 
        {M_{F'}}  {m_{\mu}}  {x_2}}
       {{ {\Delta_6}}^2} \right) \nonumber
\end{split}
\\
\begin{split}
X_7 &= 24 {x_1} \left( \frac{2  f_{7b} 
        {M_F}}{ {\Delta_7}} + 
    \frac{2  f_{7a}  {m_{\mu}}}
     { {\Delta_7}} + 
    \frac{f_{7d}  {m_f} 
        {M_F}  {m_{\mu}}  {x_2}}{
         { {\Delta_7}}^2} \right) \nonumber
\end{split}
\\
\begin{split}
X_8 &= 24 {x_1} \left( \frac{2  f_{8c} 
        {M_F}}{ {\Delta_8}} + 
    \frac{2  f_{8a}  {m_{\mu}}}
     { {\Delta_8}} + 
    \frac{f_{8d}  {m_f} 
        {M_F}  {m_{\mu}}  {x_2}}{
         { {\Delta_8}}^2} \right) \nonumber
\end{split}
\end{align}

 The $\Delta_i$'s are given by,
\bea
\Delta_1 &=& -  { {m_V}}^2\, {x_2}  - 
  { {M_F}}^2\, (  {x_1} + {x_3}  )  - 
  \Lambda^2\, ( 1 -  {x_1} -  {x_2} - {x_3}  ) \nonumber \\
\Delta_2 &=&  -  { {M_F}}^2\, {x_2} - 
  { {m_V}}^2\, (  {x_1} + {x_3}  )  - 
  \Lambda^2\, ( 1 -  {x_1} -  {x_2} - {x_3}  ) \nonumber \\
\Delta_3 &=&  -  \Lambda^2\, {x_1}   - 
  { {m_V}}^2\, {x_3} - 
  { {M_F}}^2\, ( 1 -  {x_1} - {x_2} -  {x_3}  ) \nonumber \\
\Delta_4 &=&  -  \Lambda^2\, {x_1}   - 
         {{M_F}}^2\, {x_2} - { {m_V}}^2\, {x_3} \nonumber \\
\Delta_5 &=& \Delta_6 =  -  \Lambda^2\, {x_1}   - 
  { {M_{F'}}}^2\, {x_2} - 
  { {m_V}}^2\, ( 1 -  {x_1} -  {x_2}  ) \nonumber \\
\Delta_7 &=& \Delta_8 =  -  \Lambda^2\, {x_1}   - 
  { {m_V}}^2\, ( 1 - {x_1} -  {x_2}  ) \nonumber 
\eea
The $f$'s are defined as 
\begin{xalignat}{2}
f_{1a} &= \frac{1}{32} \left((D^{L}_{V\mu^{*}\mu})^2 
                           + (D^{R}_{V\mu^{*}\mu})^2\right) &
f_{1b} &= \frac{1}{16} D^{L}_{V\mu^{*}\mu} D^{R}_{V\mu^{*}\mu} \nonumber\\
f_{2a} &= -\frac{1}{32} \left((D^{L}_{V\nu_{\mu}^{*}\mu})^2 
                            + (D^{R}_{V\nu_{\mu}^{*}\mu})^2 \right) &
f_{2b} &= \frac{1}{16} D^{L}_{V\nu_{\mu}^{*}\mu} 
                       D^{R}_{V\nu_{\mu}^{*}\mu}\nonumber\\
f_{3a} &= \frac{1}{32e}\left(
           D^{L}_{VF\mu}  D^{L}_{\gamma Ff} F^{L}_{Vf\mu} 
          + L\leftrightarrow R 
                       \right) & 
f_{3b} &= \frac{1}{32e}\left(
           D^{L}_{VF\mu}  D^{L}_{\gamma Ff} F^{R}_{Vf\mu} 
          + L\leftrightarrow R 
                       \right)
           \nonumber\\
f_{3c} &= \frac{1}{32e}\left(
           D^{L}_{VF\mu}  D^{R}_{\gamma Ff} F^{R}_{Vf\mu} 
          + L\leftrightarrow R 
                       \right) & 
f_{3d} &= \frac{1}{32e}\left(
           D^{L}_{VF\mu}  D^{R}_{\gamma Ff} F^{L}_{Vf\mu} 
          + L\leftrightarrow R 
                       \right)
           \nonumber\\
f_{4a} &= \frac{1}{32e}\left(
           F^{R}_{Vf\mu}  D^{R}_{\gamma Ff} D^{R}_{VF\mu} 
          + L\leftrightarrow R 
                       \right) & 
f_{4b} &= \frac{1}{32e}\left(
           F^{R}_{Vf\mu}  D^{R}_{\gamma Ff} D^{L}_{VF\mu} 
          + L\leftrightarrow R 
                       \right)
           \nonumber\\
f_{4c} &= \frac{1}{32e}\left(
           F^{R}_{Vf\mu}  D^{L}_{\gamma Ff} D^{L}_{VF\mu} 
          + L\leftrightarrow R 
                       \right) & 
f_{4d} &= \frac{1}{32e}\left(
           F^{R}_{Vf\mu}  D^{L}_{\gamma Ff} D^{R}_{VF\mu} 
          + L\leftrightarrow R 
                       \right)
           \nonumber\\
f_{5a} &= - \frac{C_{VF'\mu}}{32e M_F^2} 
             \left(
             D^{L}_{VF'\mu} D^{L}_{\gamma \mu^{*} \mu} 
          + L\leftrightarrow R 
                       \right) & 
f_{5b} &= -\frac{C_{VF'\mu}}{32e M_F^2} 
             \left(
             D^{L}_{VF'\mu} D^{R}_{\gamma \mu^{*} \mu} 
          + L\leftrightarrow R 
                       \right)
           \nonumber\\
f_{6a} &= -\frac{C_{VF'\mu^{*}}}{32e M_F^2} 
             \left(
             D^{L}_{V F'\mu} D^{L}_{\gamma \mu^{*} \mu} 
          + L\leftrightarrow R 
                       \right) & 
f_{6b} &= -\frac{C_{VF'\mu^{*}}}{32e M_F^2} 
             \left(
             D^{R}_{V F'\mu} D^{L}_{\gamma \mu^{*} \mu} 
          + L\leftrightarrow R 
                       \right)
           \nonumber\\
f_{7a} &= -\frac{1}{32e M_F^2} 
             \left(
             F^{R}_{Vf\mu} D^{R}_{V\mu^{*}f} D^{R}_{\gamma \mu^{*} \mu} 
          + L\leftrightarrow R 
                       \right) & 
f_{7b} &= -\frac{1}{32e M_F^2} 
             \left(
             F^{R}_{Vf\mu} D^{R}_{V\mu^{*}f} D^{L}_{\gamma \mu^{*} \mu} 
          + L\leftrightarrow R 
                       \right) 
           \nonumber\\
f_{7c} &= -\frac{1}{32e M_F^2} 
             \left(
             F^{R}_{Vf\mu} D^{L}_{V\mu^{*}f} D^{L}_{\gamma \mu^{*} \mu} 
          + L\leftrightarrow R 
                       \right) & 
f_{7d} &= -\frac{1}{32e M_F^2} 
             \left(
             F^{R}_{Vf\mu} D^{L}_{V\mu^{*}f} D^{R}_{\gamma \mu^{*} \mu} 
          + L\leftrightarrow R 
                       \right) 
           \nonumber\\
f_{8a} &= -\frac{1}{32e M_F^2} 
             \left(
             F^{L}_{Vf\mu} D^{L}_{V\mu^{*}f} D^{L}_{\gamma \mu^{*} \mu} 
          + L\leftrightarrow R 
                       \right) & 
f_{8b} &= -\frac{1}{32e M_F^2} 
             \left(
             F^{R}_{Vf\mu} D^{L}_{V\mu^{*}f} D^{L}_{\gamma \mu^{*} \mu} 
          + L\leftrightarrow R 
                       \right) 
           \nonumber\\
f_{8c} &= -\frac{1}{32e M_F^2} 
             \left(
             F^{R}_{Vf\mu} D^{R}_{V\mu^{*}f} D^{L}_{\gamma \mu^{*} \mu} 
          + L\leftrightarrow R 
                       \right) & 
f_{8d} &= -\frac{1}{32e M_F^2} 
             \left(
             F^{L}_{Vf\mu} D^{R}_{V\mu^{*}f} D^{L}_{\gamma \mu^{*} \mu} 
          + L\leftrightarrow R 
                       \right) 
           \nonumber
\end{xalignat}
%

%
%

\end{document}